\documentclass{aa}
\usepackage[utf8]{inputenc}
\usepackage[varg]{txfonts}
\usepackage{graphicx}
\usepackage{color}
\usepackage[normalem]{ulem}
\PassOptionsToPackage{hyphens}{url}
\begin{document}
\newcommand\redsout{\bgroup\markoverwith{\textcolor{red}{\rule[0.5ex]{2pt}{0.4pt}}}\ULon}

\title{Inference of the cosmic rest-frame from supernovae Ia}
\author{Nick Horstmann \and Yannic Pietschke \and Dominik J. Schwarz }

\institute{Fakult\"at f\"ur Physik, Universit\"at Bielefeld, 
Postfach 100131, 33501 Bielefeld, Germany \\ 
\email{nhorstmann, ypietschke, dschwarz at physik.uni-bielefeld.de}}

\date{Received <date> / Accepted <date>}

\abstract{We determine the proper motion of the Solar system 
from the Pantheon sample of supernovae (SNe) of type Ia. 
The posterior
distribution of the Solar system proper velocity, its direction 
and relevant cosmological parameters are obtained based on the 
observed distance moduli, heliocentric redshifts, and positions 
of SNe by means of a Markov Chain Monte Carlo method. 
We account for the unknown peculiar motion of SNe 
by including their expected covariance from 
linear theory. We find that the Solar system moves 
with $v_o = 249 \pm 51$ km/s towards $RA = 166 \pm 16$ deg, 
$Dec = 10 \pm 19$ deg (J2000) (all at 68\% C.L.). The direction 
of motion agrees with the direction of the dipole observed in 
the cosmic microwave background (CMB) 
($RA = 166$ deg, $Dec = -7$ deg). The inferred velocity is 
$2.4 \sigma$ smaller than the value inferred 
from a purely kinematic interpretation of the CMB dipole 
($370$ km/s). 
Assuming a flat $\Lambda$ cold dark matter model, we find no 
degeneracy of Solar proper motion with other cosmological parameters. 
The dimensionless matter density, $\Omega_M = 0.305 \pm 0.022$, is 
in excellent agreement with CMB measurements. We also find no
degeneracy of the Solar proper motion with the SN calibration nuisance 
parameter. Bulk flows might be able to explain why the solar motion appears to be slower w.r.t.~nearby SNe. We conclude that a larger sample of SNe, distributed over wide areas of sky and a broad range in redshift, will
allow an 
independent and robust test of the kinematic nature of the CMB dipole.}


\maketitle
\section{Introduction}

The existence of a cosmic rest-frame follows from  
the cosmological principle, which states that the Universe is 
spatially homogeneous and isotropic (at large enough scales or 
in a statistical sense). Any free-falling, non-rotating observer 
will asymptotically (for a Universe that exists forever) come to 
rest with respect to that frame,
i.e.~all peculiar velocities become asymptotically small.\footnote{See e.g.~\cite{2006MNRAS.373..382B} for a detailed discussion. 
Here we refer to their definition 3, which is equivalent to their definition 7 (the observer measures a vanishing CMB dipole) and implies their definition 1 (the change of comoving distance decays asymptotically).} This is consistent with the observed 
smallness (i.e.~$v \ll c$) of the peculiar motions.

In this work, we use the Pantheon sample of supernovae (SN) 
of type Ia \citep{Scolnic2018} to measure the peculiar motion of the 
Solar system w.r.t.\ a cosmic rest frame 
defined by the SN sample itself and compare it with the 
peculiar motion inferred from observations of the cosmic microwave 
background (CMB) \citep{Planck2013XXVII,Saha2021}. 
We use the measured heliocentric redshifts of the 
Pantheon SNe, which have been neatly compiled in a catalogue by 
\citet{Steinhardt2020}. 

Shortly after the discovery of the CMB, which is dominated by a thermal
monopole at $T_0 = 2.7255 \pm 0.0006$ K \citep{Fixsen2009}, the frame 
of this cosmic heat bath has been identified with the cosmic rest frame
\citep{Stewart1967,PeeblesWilkinson1968}. A peculiar motion of the 
Solar system gives rise the so-called Solar dipole, 
which is modulated by the motion of Earth around the Sun, the 
so-called annual dipole. 
An observer moving with velocity $\mathbf{v}$ relative to a heat bath 
with a temperature $\bar T$ observes the temperature
\begin{equation}
T(\mathbf{n}) = \bar T \frac{1 - \mathbf{v}^2/c^2}{1 - \mathbf{v}\cdot\mathbf{n}/c} 
    = T_0 + T_1 \cos \theta + {\cal O}[(v/c)^2],
\end{equation}
towards a direction given by the unit vector $\mathbf{n}$, 
where $\cos \theta = \mathbf{v} \cdot \mathbf{n}/v$, 
$T_0 = \bar T + {\cal O}[(v/c)^2]$ and 
$T_1 = \bar T v/c + {\cal O}[(v/c)^3]$, 
with $v$ denoting the speed of the observer 
and $c$ the speed of light. 

The CMB dipole, as measured w.r.t.\ the barycentre of the 
Solar system, was determined most precisely by
\citet{PlanckI}, with a dipole 
amplitude of $T_1 = 3.36208 \pm 0.00099$ mK and pointing 
towards 
$RA = (167.942 \pm 0.007)$ deg, 
$Dec = (- 6.944 \pm 0.007)$ deg (J2000).
It is commonly assumed that the frame defined by the observed CMB dipole 
can be identified with the cosmic rest frame, which implies that the 
Solar system moves at $v/c = (1.23356 \pm 0.00045) \times 10^{-3}$, 
where the uncertainty is dominated by the uncertainty of $T_0$.

However, the kinematic origin of the CMB dipole is a hypothesis that must be 
tested, and there are good reasons to believe that subdominant contributions
to the CMB dipole are due to non-kinematic effects. If primordial
fluctuations of space-time and matter are seeded via quantum fluctuations
during cosmological inflation, we would expect contributions to the 
dipole from the Sachs-Wolfe effect, the proper motion 
of the last scattering surface, and an integrated
Sachs-Wolfe effect in the early and late Universe. Extrapolating the Planck 
best-fit to a cosmological constant and cold dark matter model
($\Lambda$CDM model), we expect the dipolar angular band power 
to be ${\cal D}_1 \approx (33 \mu\mathrm{K})^2$.
\footnote{We use the cosmological parameters of the best-fit flat-$\Lambda$CDM model from \cite{PlanckVI} and {\tt CAMB} \citep{camb} to obtain the expected primordial CMB dipole moment $C_1$, which gives the angular band power ${\cal D}_1 = C_1/\pi$.} This corresponds to a 
primordial contribution to the dipole amplitude $T_1$ of the order of 
$\delta T_1 \approx 0.05$ mK.
Recently, an upper limit on the 
intrinsic CMB dipole was presented in \cite{Ferreira2021}. 
It is possible to use higher multipole moments of the CMB to infer the 
Solar motion \citep{Challinor2002}, as it induces a coupling of neighbouring 
multipole moments due to Doppler boosting and aberration. In contrast 
to the direct measurement of the CMB dipole, the precision of that 
measurement is limited by the number of hot and cold spots of the CMB 
that can be resolved with high signal to noise.
\citet{Planck2013XXVII,Planck2015XVI} showed 
that the kinematic dipole assumption is consistent with the anisotropies 
of the CMB at angular scales below one degree. 
Most recently \citet{Saha2021} refined that analysis and found 
$v/c = (0.996 \pm 0.219) \times 10^{-3}$ (about one $\sigma$ smaller than the value inferred from the CMB dipole), with a large directional 
uncertainty of tens of degrees. Measuring the CMB at more frequency bands would in principle allow to improve the accuracy, but has to await the next CMB space mission.

While a comparison of the CMB dipole with the dipolar modulation of 
small-scale features of the CMB allows us, at least in principle, to 
disentangle a kinematic and a primordial intrinsic dipole, 
a measurement of the dipolar features of the Universe at different 
redshift would additionally allow us to test the question 
if a cosmic (i.e.\ a universal) rest frame exists at all, which would
also test the cosmological principle \citep{Schwarz2015}.

A simple method to measure the rest frame in the late Universe is
based on number counts of distant objects \citep{EllisBaldwin1984}. 
It can be applied on radio surveys which cover large fractions of 
the sky, probe objects at high redshift, and do not suffer from 
dust extinction. Several radio surveys from different instruments 
were analysed \citep{Blake2002,Singal2011,Rubart2013,Tiwari2015,Singal2019,Siewert2021}, 
spanning a full decade in radio frequency. All 
radio source count dipoles agree within errors with the direction of 
the CMB dipole, but strongly disagree with the amplitude of the CMB dipole
when based on the assumption of a kinematic origin of this matter dipole.
The inferred Solar velocity would be between a factor of four to ten 
larger than inferred from the CMB, increasing with decreasing radio 
survey frequency. If confirmed, a chromatic radio dipole  
excludes a simple kinematic interpretation \citep{Siewert2021}. 

This excess matter dipole was confirmed by a similar analysis 
of quasars from infrared data \citep{Secrest2021}. Again, the quasar 
dipole points towards the CMB dipole direction, but shows a 
significant excess of the dipole amplitude. This is consistent with 
the findings at the higher radio frequencies, which also probe a
similar sample of objects, i.e.\ radio sources that are mostly active 
galactic nuclei. 

These findings call for a closer investigation of the question of the 
cosmic rest frame, which is assumed to be given by the Solar dipole 
in almost all studies of observational cosmology, including the 
Hubble diagram. 

The observed Hubble tension between global and local 
measurements of the Hubble rate, see e.g.\ between \citet{PlanckVI} 
and \citet{Riess2021}, may actually be a hint towards a fundamental 
issue. This also raises the question of a direct test of the 
isotropy of the Hubble diagram, see e.g.\ 
\citep{McClure2007,Schwarz2007,Antoniou2010,Kalus2013}, 
which is of course also frame dependent. 
Recently, there have been contradicting claims in the literature 
on how anisotropic \citep{Colin2019,Singal2021,Krishnan2021} or 
isotropic \citep{Hu2020,Rahman2021} the expansion actually is.

Tightly related is also the discussion on large scale bulk flows, 
see e.g.\ \citet{Colin2011,Feindt2013}. To our knowledge, recent works in 
that context are based on the kinematic interpretation of the CMB dipole
and to our surprise, only little effort has been invested to 
measure the Solar motion by means of SNe out to large cosmological 
scales \citep{Singal2021}. 
A crucial issue in that context are peculiar velocities of 
SNe and their host galaxies, as well as the consistent treatment of the 
SNe redshifts, see e.g.\ \citet{Davis2019,Peterson2021}.

This work is structured as follows. In Sec.~\ref{sec:2} we describe and
summarize the effects of peculiar motion on the Hubble diagram. 
Besides the Doppler effect on the redshift, the distance modulus is 
affected as well. While typically SN samples are published in the 
CMB frame with regard to the reported redshifts, the reported 
distance moduli are the heliocentric ones. The SN data used in this 
analysis is described in Sec.~\ref{sec:3}, with some additional 
details in App.~\ref{appendix:DataErrors}. Section \ref{sec:4} 
describes the Markov Chain Monte Carlo method that we use to obtain 
the posterior distributions and a suite of consistency checks. More 
details on those checks are provided in App.~\ref{appendix:SubSamples}. Our 
results are presented in Sec.~\ref{sec:5} and we conclude in 
Sec.~\ref{sec:6}. 

\section{Distance modulus and the observers motion \label{sec:2}}

Let us start from an idealised measurement of luminosity distance between a 
source and an observer, both being at rest 
with respect to a homogeneous and isotropic space-time, $\overline{d}_L(\overline{z})$. 
We call that frame the cosmic rest-frame. 
We then introduce peculiar motions of the observer $\vec{v}_o$ and emitter $\vec{v}_e$, 
following the discussion of \citet{Hui2006}. The dependence of the luminosity distance 
on relativistic Doppler shift and aberration was first studied in the seminal work 
of \citet{Sasaki1987}. In the following, overlined symbols denote quantities without 
peculiar motions.

The (line-of-sight) comoving (with the cosmic rest frame) distance is defined as
\begin{align}
\chi(\overline{z})= c \int_0^{\overline{z}}\frac{dz^\prime}{H(z^\prime)} ,
\end{align}
where the Hubble rate $H(z)$ for the $\Lambda$ cold dark matter model is given by
\begin{align}
H(z)=H_0\sqrt{\Omega_M (1+z)^3 + \Omega_K(1+z)^2 +(1-\Omega_M-\Omega_K)},
\end{align}
with $\Omega_\Lambda = 1 - \Omega_M - \Omega_K$ and $\Omega_\Lambda, \Omega_M$, 
and $\Omega_K$, denoting the dimensionless energy densities of cosmological 
constant, matter, and curvature, respectively, and $H_0 = H(0)$. The bolometric luminosity 
distance is
\begin{align}
\overline{d}_L(\overline{z})=\frac{(1+\overline{z})}{\sqrt{\Omega_K}} \frac{c}{H_0} \sinh{\left[  \sqrt{\Omega_K} (H_0/c) \chi(\overline{z})\right]}.
\end{align}
It is related to the angular diameter distance via the Etherington distance duality relation
\begin{align}
\label{ddr}
\overline{d}_L(\overline{z})= \overline{d}_A(\overline{z})(1+\overline{z})^2,
\end{align}
which holds for arbitrary space-time geometries, as long as light rays propagate in vacuum 
and as long as the limit of geometric optics applies \citep{Etherington1933, Schulze-Koops2017}.

When allowing for peculiar motion, the first effect to take into account is the Doppler shift 
experienced by light due to relative motion of source and observer. The redshift $z$ 
measured in the heliocentric frame, is connected to the redshift $\overline{z}$ in the cosmic 
rest-frame, or comoving frame, via
\begin{align}
\frac{1+z}{1+\overline{z}}=1+z_{\mathrm{Doppler}}=\gamma (1+(\vec{v}_e-\vec{v}_o)\cdot \vec{n}/c) ,
\end{align}
where $\gamma=(1-(\vec{v}_e-\vec{v}_o)^2/c^2)^{-1/2}$ is the Lorentz factor. Here, $\vec{n}$ is 
the unit vector pointing from the observer to the emitter.
Expanding up to first order in velocities, we obtain
\begin{align}
1+z=(1+\overline{z})\left[1+ (v_e/c) \cos \phi - (v_o/c) \cos \theta\right]+\mathcal{O}(v^2) . \label{eq: z}
\end{align}
$\phi$ and $\theta$ denote the observed 
angles between the line of sight and the peculiar motion of emitter and observer, respectively. 
$\mathcal{O}(v^2)$ indicates second order terms in $\vec{v}_e$ and $\vec{v}_o$. Aberration of 
the angles can be ignored at linear order, because the rest-frame angle $\overline \theta$ 
transforms as $\cos \overline \theta = \cos \theta + \mathcal{O}(v)$ and thus, effects on the 
redshift are of order $\mathcal{O}(v^2)$. The same holds true for $\overline \phi$.

However, relativistic aberration gives rise to a first order effect in the angular diameter distance
\begin{align}
d_A=\sqrt{\frac{\delta A_e}{\delta \Omega_o}} , 
\end{align}
where $\delta A_e$ is the proper area of the emitter and $\delta\Omega_o$ is the solid angle 
at the observer subtended by light rays from the emitter. When considering relative motion, 
the latter will change. By performing Lorentz boosts, one can see that, the solid angle element 
transforms as
\begin{align}
d\Omega = \overline{d\Omega} \frac{1-v_o^2/c^2}{\left[1+ (v_o/c) \cos \overline \theta\right]^2}
= \overline{d\Omega} \left[1 - 2 (v_o/c) \cos \theta \right]  + \mathcal{O}(v^2). 
\end{align}
The angular diameter distance is then given by
\begin{align}
d_A(z)=\overline{d}_A(\overline{z})\left[1+(v_o/c) \cos \theta \right] +\mathcal{O}(v^2) .
\end{align}
Thereby, from (\ref{ddr}), the luminosity distance can be written as
\begin{align}
d_L(z)= \overline{d}_L(\overline{z})
\left[1 + 2 (v_e/c) \cos \phi - (v_o/c) \cos \theta \right] +\mathcal{O}(v^2).
\end{align}

It remains to link the bolometric luminosity distance to the bolometric distance modulus, 
\begin{align}
\mu = m_B - M = 5 \log_{10}\left[ \frac{d_L(z)}{\mathrm{Mpc}}\right]+25,
\end{align}
which is inferred from the analysis of SN
lightcurves, up to their absolute bolometric 
magnitude, which is treated as a nuisance parameter in the cosmological analysis (except when measuring $H_0$).
We must also express $\overline{z}$ as a function of the measured redshift,
\begin{align}
    \overline{z}(z)=z + (1+z)\left[(v_o/c) \cos \theta - (v_e/c) \cos \phi \right] + \mathcal{O}(v^2). 
    \label{eq: redshift}
\end{align}

Then the fitting function, at linear order in peculiar velocities, reads
\begin{align}
    \mu(z)= &5 \log_{10}\left[ \frac{\overline{d}_L \left(\overline{z}(z)\right)}{\mathrm{Mpc}}\right] + 25 \notag\\
    + &5 \log_{10}\left[1+2(v_e/c) \cos \phi - (v_o/c) \cos \theta\right]. 
    \label{eq: mu}
\end{align}

While we can hope to measure the velocity of the Solar system accurately, it is much harder to 
measure the peculiar motion of extragalactic objects, especially their proper motion is usually 
not detectable with current observational precision. Thus $\mathbf{v}_e$ has to be inferred 
from incomplete and noisy data, e.g. see \citet{Carrick2015}. An essential step in the reconstruction 
of peculiar velocities is that one must assume that the cosmic rest-frame is known. 

In this work we wish to determine the cosmic rest-frame based on a sample of SNe of type Ia
and thus, it would be inconsistent to assume 
peculiar velocities that make already an assumption with respect to the quantities that we 
wish to measure. We therefore must refrain 
from applying peculiar velocity corrections 
for extragalactic objects, but rather take the corresponding uncertainties into account by 
including them in the covariance matrix 
in an appropriate way. 

We do so following \citet{Huterer2017}. The covariance matrix can be written as (the indices 
denote individual supernovae) 
\begin{align}
\label{covariance_matrix}
C_{ij} =  N_{ij} + S_{ij} + V_{ij},
\end{align}
where $N_{ij}$ is the observational contribution to the covariance and includes uncertainties 
from photometry and lightcurve fitting, while  $S_{ij}$ represents the additional covariances 
between the different surveys within the complete data catalog, as described for the 
Pantheon SN catalogue in \citet{Scolnic2018}. $V_{ij}$ denotes the expected 
variances from peculiar motion and covariances from bulk flows, as predicted in linear 
perturbation theory. It reads, 
\begin{align}
\label{eq:pecVelCov}
    V_{ij} = \langle \Delta m_i \Delta m_j \rangle = \left( \frac{5}{\ln 10}\right) ^2
    \frac{(1+\overline{z}_i)^2}{H(\overline{z}_i)d_L(\overline{z}_i)}
    \frac{(1+\overline{z}_j)^2}{H(\overline{z}_j)d_L(\overline{z}_j)}\xi _{ij},
\end{align}
where 
\begin{align}
    \xi _{ij} = \langle (\mathbf v_i\cdot \mathbf{n}_i)(\mathbf v_j \cdot \mathbf{n}_j)\rangle 
    = &\frac{dD_i}{d\tau}\frac{dD_j}{d\tau}\int \frac{dk}{2\pi^2}P(k,\overline{z}=0)\, \times \nonumber \\
    \sum ^{l_{max}}_{l = 0}(2l+1) &j^\prime _l(k\chi _i)
    j^\prime _l(k\chi _j)(P_l(\mathbf{n}_i\cdot \mathbf{n}_j)-\delta _{l0}). 
\end{align}
Above, $P(k,z=0)$ denotes the linear power spectrum today, $D_i$ is the linear growth factor, 
$\tau$ denotes conformal time, 
$j_l$ are spherical Bessel functions and the 
prime denotes a derivative w.r.t.\ their argument.
Finally, $P_l$ are Legendre polynomials. 

It has been criticised \citep{Mohayaee2020} that this approach would
not account for the fact that we observe the 
Universe from within a galaxy and that neither 
the Milky Way nor the Solar system are 
comoving with the cosmic rest-frame, while 
the derivation of the structure covariance (\ref{eq:pecVelCov}) 
assumes that the observer is comoving and sits 
at a random location in the Universe, 
i.e.\ at a spot without a galaxy. This is 
certainly true, and this aspect
could be improved by looking at constrained 
three-point correlations instead. However, we 
expect that the effect of that correction is 
small as we are not including very nearby SNe. 
The smallest SN distance in the considered
sample is about 40 Mpc ($z \sim 0.01$) and 
the majority of SNe is at much larger
distances. At distances above 40 Mpc, 
linear perturbation theory is 
a very reasonable approximation and 
non-vanishing three-point correlations are 
not expected for Gaussian initial conditions 
in the linear regime.

\begin{table}
\caption{Surveys used for the Pantheon catalogue. 
}
\label{tb: PantheonSurveys}
\begin{tabular}{l|c|c|c|c}
    Survey & SNe & SNe  & redshift & redshift  \\
    name & all  & with hostz & mean & range \\
    \hline
    SNLS & 236 & 175 & 0.638 & 0.124 - 1.060 \\
    SDSS & 380 & 250 & 0.186 & 0.012 - 0.403 \\
    PS1 & 279 & 132 & 0.288 & 0.026 - 0.630 \\
    CfA 1-4 & 102 & 102 & 0.029 & 0.009 - 0.075 \\
    CSP & 25 & 25 & 0.028 & 0.011 - 0.058\\
    HST & 26 & 18 & 1.278 & 0.735 - 2.260 \\
    \hline
    Pantheon & &  & & \\ 
    all & 1048 & 702 & 0.323 & 0.009 - 2.260 \\
    hostz & 702 & 702 & 0.301 & 0.009 - 2.260
\end{tabular}{}
\tablefoot{The columns show the total number of SNe, the number of SNe with 
known host redshift, mean heliocentric redshift, and heliocentic 
redshift range, according to \citet{Steinhardt2020}.}
\end{table} 

\section{The supernova sample \label{sec:3}}

\subsection{Pantheon catalog}

We use the Pantheon sample consisting of 
1048 SNe of type Ia, originally compiled by \citet{Scolnic2018}. 
The Pantheon sample is a compilation of 
SN data from the surveys specified in Tab.~\ref{tb: PantheonSurveys}. Those are the 
Supernova Legacy Survey (SNLS) \citep{Guy2010}, 
the Sloan Digital Sky Survey (SDSS) \citep{Smith2012, Sako2018}, 
Pan-STARS1 (PS1) \citep{Scolnic2018}, CfA1 -- CfA4 
\citep{Riess1999,Jha2006,Hicken2009a,Hicken2009b,Hicken2012}
the Carnegie Supernova Project (CSP) \citep{Contreras2010} and various surveys using the 
Hubble Space Telescope (HST), namely CANDLES/ CLASH 
\citep{Rodney2014,Graur2014,Riess2018}, GOODS \citep{Riess2007} and SCP \citep{Suzuki2012}.

\begin{figure}
    \centering
    \includegraphics[width=0.49\textwidth]{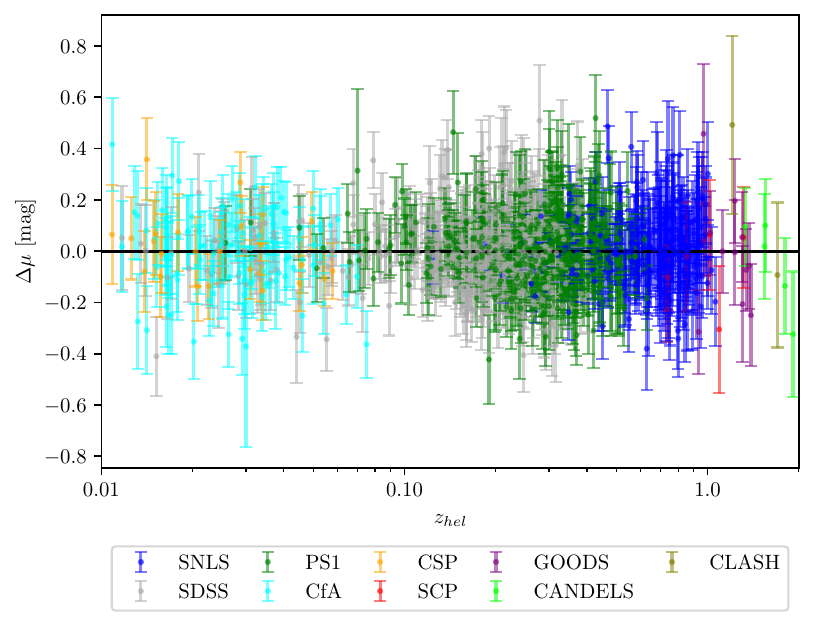}
    \caption{Residual of Hubble diagram as a function of the observed heliocentric redshift. 
    The different colors indicate the redshift
    distributions of the individual SN surveys included in the Pantheon sample.\label{fig1}}
\end{figure}

\begin{figure}
    \centering
    \includegraphics[width=0.49\textwidth]
    {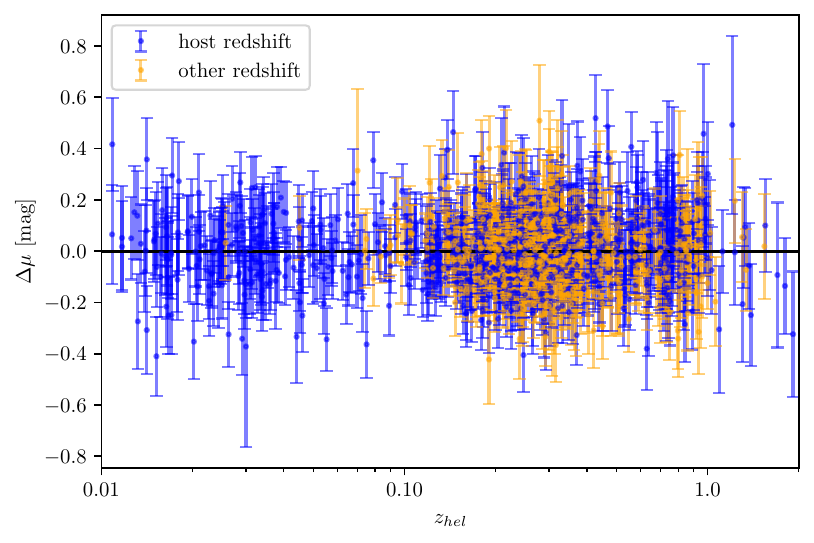}
    \caption{Residual of Hubble diagram as a function of the observed heliocentric redshift 
    indicating SNe with host redshift measurements 
    and with other redshift measurements. \label{fig2}}
\end{figure}

\subsection{SN redshift}

It has been pointed out by \citet{Davis2019} that the 
redshifts of SNe should be considered with great care and 
commonly made mistakes in the handling of redshift estimates 
could lead to false cosmological conclusions. After 
\citet{Rameez2019} noticed several inconsistencies 
between redshifts quoted in the Pantheon \citep{Scolnic2018} 
and JLA \citep{jla} SN catalogues,  
\citet{Steinhardt2020} identified the cause of several of these 
inconsistencies regarding the redshift values quoted in the 
Pantheon catalogue and published an 
improved catalogue.\footnote{\url{iopscience.iop.org/article/10.3847/1538-4357/abb140}}
Here we make use of this improved catalogue after implementing 
additional corrections related to the positions of several SNe from the 
HST surveys as discussed  in App.~\ref{appendix:DataErrors}. 
The improved Pantheon catalogue \citep{Steinhardt2020} provides detailed information on 
each SN: the SN identifier, the original survey, 
the type of redshift measurement (SN, host, both), as well as values for $z_\mathrm{CMB}$, $z_\mathrm{hel}$, 
$\mu $, $RA$, $Dec$ and the measurement uncertainties of $z_\mathrm{CMB}$ and 
$\mu $. The quoted redshift uncertainties 
are not fully self-consistent, as pointed out by \citet{Steinhardt2020}. For example, some 
uncertainties are given as $0$, because the original survey reported no uncertainties for these. 
For other surveys only a general uncertainty is given and not an individual 
one for each SN. For this reason, we choose to neglect redshift uncertainties in our analysis.

As we wish to measure the proper motion of the 
Solar system, it is important to distinguish between the cosmological redshift, 
$\Bar{z} = z_\mathrm{CMB}$, and the observed heliocentric redshift, $z = z_\mathrm{hel}$, which includes 
effects of peculiar motion (here we ignore other effects, e.g.\ gravitational redshift). 
The connection is given by \eqref{eq: redshift}. 
One of the aspects pointed out by \citet{Steinhardt2020} in their equation (A.1) is the fact that \citet{Scolnic2018} 
omit the term $z (v_o/c)\cos{\theta}$, giving rise to errors for high redshifts. 

In a first step, that just serves for illustrative purposes, we fit (\ref{eq: mu}) 
to the full redshift-improved Pantheon sample, 
without including peculiar velocities of 
the emitters and without taking the full covariance matrix $C_{ij}$ into account
(just the variances of $\mu$ as quoted in the catalogue). 
To illustrate the 
redshift distribution of the various 
subsamples of the Pantheon SN sample 
we show the residuals to a fit to the Hubble 
diagram in Fig.\ \ref{fig1}. We varied 
$H_0$ (treated as a nuisance parameter), $\Omega_M$ and $v_o$, for a flat 
$\Lambda$CDM model and 
fixed the direction of the peculiar 
of motion of the Solar system to the 
CMB dipole direction. The fit is obtained with the help of {\tt lmfit}\footnote{\url{lmfit.github.io/lmfit-py/}} \citep{lmfit} and the corresponding figures are produced with {\tt matplotlib}\footnote{\url{matplotlib.org/}} \citep{matplotlib}. We find a 
good quality of fit with $\chi^2/$dof $= 1.0001$ and best fit values: 
$\Omega _M = 0.30 \pm 0.01$ and $v_o / c = 0.0008 \pm 0.0001$.
\citet{Steinhardt2020} also provide the type of redshift measurement for each SN. The main distinction is between measurements that are 
inferred from the SN and redshifts inferred from the spectra of their respective host galaxy (hostz). \citet{Steinhardt2020} state, that the hostz values are more reliable and recommend using hostz measurements in future surveys. They also report different fit results between the two methods for cosmological parameters. For this reason, we looked at the hostz subset separately. We find that there is no substantial difference in the value of the inferred $v_o$, just an increase 
in the uncertainty, consistent with the reduced number of data points ($\chi^2/$dof = 1.013). 
The corresponding residuals to the Hubble diagram are shown in Fig.\ \ref{fig2}. We therefore use the complete Pantheon sample for our analysis below, 
and use the subsample of SNe with known host galaxy redshifts for a consistency check. 

To estimate the peculiar velocities of SNe with $z_\mathrm{hel} < 0.1$, included in the inference of  $z_\mathrm{CMB}$ by \citet{Steinhardt2020}, we used a tool published in conjunction with \citet{Carrick2015}. This tool requires the cosmological parameters $H_0$ and $\Omega _M$ 
as well as for each supernova $z_\mathrm{CMB}$, 
$RA$ and $Dec$ as input parameters. For $RA$ and $Dec$ we directly use the values given by \citet{Steinhardt2020}. $z_\mathrm{CMB}$ is calculated from the heliocentric values given by \citet{Steinhardt2020} according to \eqref{eq: redshift}, without 
$v_e$, using the solar velocity we got from our first analysis.

When assuming a velocity of $v_o/c = 0.00123$ (\citet{PlanckI}), we recover the values of $\Bar{z}$ given by \citet{Steinhardt2020} 
for $z>0.1$ within a numerical uncertainty of about $10^{-6}$.  For smaller redshifts the difference to the catalogued $\Bar{z}$ is significantly larger, both when taking peculiar velocities into account (as described above) and when omitting them.

However, reconstructing peculiar velocities from density fields requires the knowledge of the cosmic rest-frame. 
Because the goal of our work is to find 
this rest-frame, peculiar velocities of the emitter are difficult to include in a consistent way. For this reason, we only 
include peculiar velocity corrections as a consistency check in our analysis, with 
the goal to get an estimate for their influence on our dipole values.

\begin{figure*}
\centering
\includegraphics[width=0.75\textwidth]
{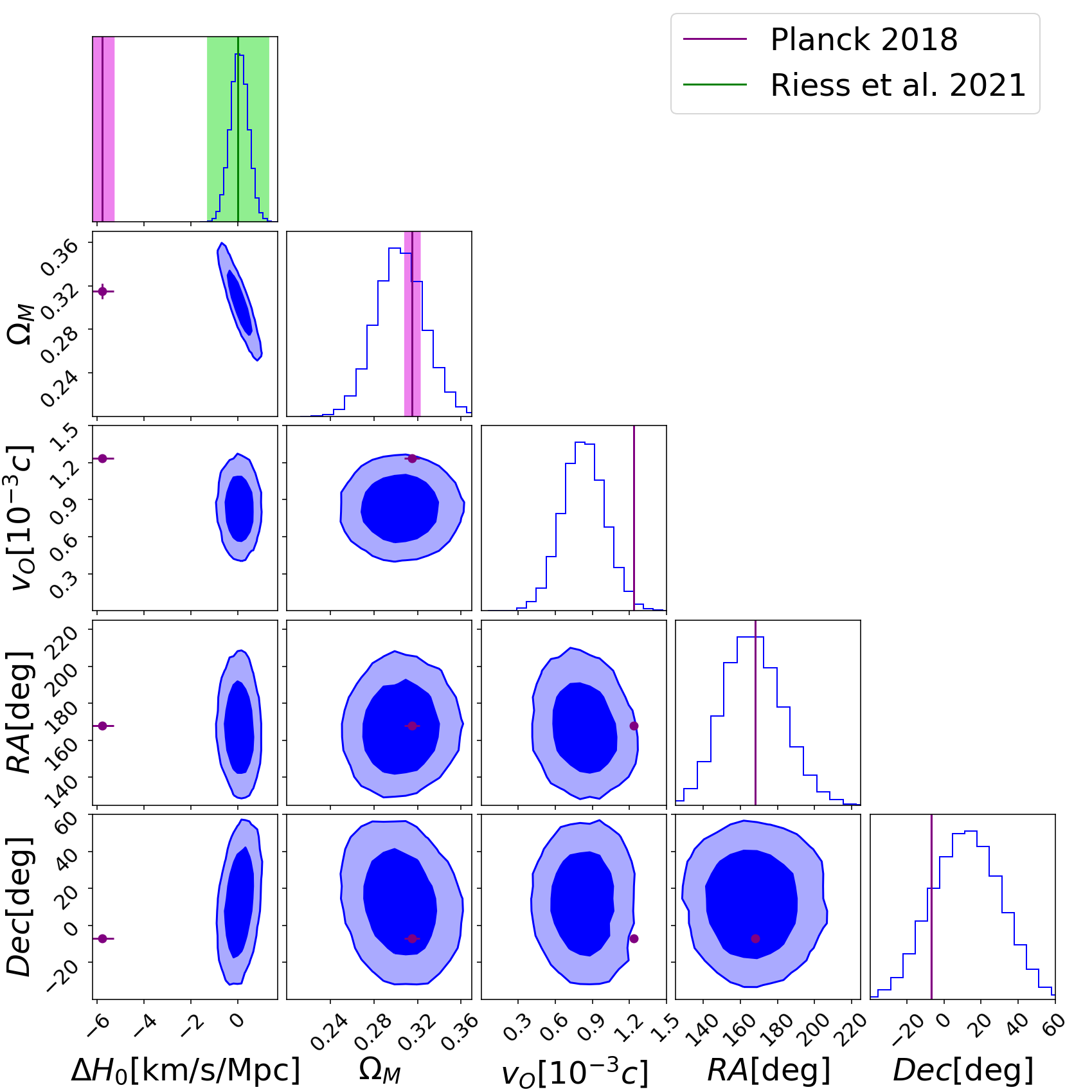}
\caption{Two-dimensional and one-dimensional posterior distributions for
four model parameters and the nuisance parameter 
$\Delta H_0$ for the full redshift-corrected Pantheon sample. 
The contours show 68\% and 95\% credibility levels. 
\label{fig:contPant}}
\end{figure*}

\citet{Steinhardt2020} adjusted the value of M in the distance moduli $\mu$ such that a fit yields a value of $70$ km/s/Mpc for $H_0$. We changed the absolute magnitude such that our fit agrees with 
$H_0 = 73.2 \ \text{km/s/Mpc}$ in line with \citet{Riess2021}, when not including any peculiar motions of the sources and fixing the Solar peculiar motion to the value inferred from the CMB \citep{PlanckI}. 

\section{Parameter fitting and consistency \label{sec:4}}

\subsection{Method}

We follow a Bayesian approach based on a Markov Chain Monte 
Carlo (MCMC) to compute the posterior probability for a set 
of five parameters ($\Delta H_0$, $\Omega _M$, $v_o$, $RA$, $Dec$),
based on the redshift-corrected Pantheon data set described above. 
The quantity $\Delta H_0 \equiv H_0 - 73.2 \ \text{km/s/Mpc}$ is treated as a nuisance parameter and use
flat priors for all parameters, allowing them to vary in broad ranges: $-73.2\ \text{km/s/Mpc} < \Delta H_0 < 46.8\ \text{km/s/Mpc}$, $0<\Omega _M < 1$, $0< v_o < 0.002c$, $0\ \text{deg} < RA < 360\ \text{deg}$, $-90\ \text{deg} < Dec < 90\ \text{deg}$, and $-0.3 < \Omega _K < 0.3$.

We use {\tt emcee} \citep{emcee2013} as our implementation of the
MCMC. Our likelihood function is, up to an irrelevant constant
containing also the determinant of the covariance matrix $C$, 
given by
\begin{align}
    \log (L) =  -\frac 12 \sum_i \sum_j \Delta \mu_i ^T (C^{-1})_{ij} \Delta \mu_j + \mbox{const},
\end{align}
where $\Delta \mu_i = \mu_\mathrm{data\ i} -
\mu_\mathrm{model\ i}$ is the distance 
modulus residual for SN with index $i$. 
The covariance matrix is $C = N + S + V$, see 
text below equation (\ref{covariance_matrix}).
$N$ is diagonal and contains the 
uncertainties in $\mu_\mathrm{data\ i}$, 
$S$ describes correlations between 
the different surveys and SNe, both are 
given by \citet{Scolnic2018}. 
$V$ accounts for correlations in peculiar velocities and is given by expression (\ref{eq:pecVelCov}), see also \citet{Huterer2017}.  Strictly speaking, $V$ 
depends on the cosmological parameters and should be recalculated in 
every iteration. 
Because this is computationally expensive, 
we fixed all required parameters to the 
best-fit values given by by the Planck 2018 results \citep{PlanckI, PlanckVI}, assuming the neglected effects to be small.

Figure \ref{fig:contPant} shows 
posterior distributions of the model parameters. The first one,
$\Delta H_0$, is included to investigate potential 
degeneracy of this nuisance parameter with the cosmological
parameters in question, but it is clear that we have fixed 
the absolute magnitude of SN Ia to an arbitrary value and thus $\Delta H_0$ must not be mistaken as a measurement of 
the Hubble rate.
The second parameter is $\Omega_M$. 
The remaining three parameters 
provide the credibility regions for the peculiar velocity of the Solar system, and its direction of motion in terms of $RA$ and $Dec$.

All contour plots in this work are created using the {\tt corner} module for Python \citep{corner2016}
and show 68\% and 95\% credibility regions. The values labeled `Planck 2018' indicate the 
Planck 2018 measurements and their 68\% credibility intervals \citep{PlanckI,PlanckVI}. The values labeled 
`Riess et al. 2021' are from \citet{Riess2021}. 

\subsection{Consistency of subsamples}

Before turning to the discussion of our main result, we analyse each SN survey of table \ref{tb: PantheonSurveys} and compare the results of these sub-samples to 
check for any inconsistencies between the different surveys. The corresponding posterior distributions and a more detailed discussion are presented in Appendix \ref{appendix:SubSamples}. 

We find that each individual SN sample from a single survey either agrees with the value $\Omega _M$ found by \cite{PlanckVI} within 1 $\sigma$ (PS1, SDSS, SNLS, HST), or has no constraining power due to a lack of high redshift SNe in these surveys (CfA, CSP). 
We also run our algorithm with different combinations of surveys and find self-consistency of the total catalogue with 
its subsamples. 

\section{Results \label{sec:5}}

Let us turn to the closer inspection of 
Fig.~\ref{fig:contPant}, which shows the one and two dimensional
posterior distributions of the five free parameters of our base 
model using the whole Pantheon data set. We assume a flat 
$\Lambda$CDM model and account for peculiar motions by
means of an additional contribution to the covariance matrix, 
see (\ref{eq:pecVelCov}). The numerical results 
of the fit are provided in table \ref{tb:contPant}. The 
proper motion of the Solar system is specified 
by the three parameters ($v_o$, $RA$, $Dec$).
The corresponding best-fit values inferred from the 
CMB are summarised from literature in table \ref{tb:contPlanck}.

\begin{table*}
\centering
\caption{Best-fit values (median) and $68$\% credible intervals.
}
    \begin{tabular}{c|c|c|c|c|c|c}
        & default & hostz & $v_e$ correction & curvature & $H_0$ fixed &  bulk motion \\
        \hline
        $\Delta H_0$ [km/s/Mpc] & $0.23^{+0.39}_{-0.37}$ & $0.58^{+0.38}_{-0.38}$ & $0.47^{+0.38}_{-0.40}$ & $0.27^{+0.45}_{-0.47}$ & 0 & $0.219_{-0.402}^{+0.373}$ \\
        $\Omega _M$ & $0.305^{+0.022}_{-0.022}$& $0.271^{+0.021}_{-0.021}$ & $0.295^{+0.023}_{-0.021}$ & $0.312^{+0.060}_{-0.064}$ & $0.312^{+0.022}_{-0.024}$ &  $0.304_{-0.021}^{+0.023}$\\
        $\Omega _K$ & 0  & 0  & 0  & $-0.018^{+0.160}_{-0.154}$ & 0 & 0\\
        $v_o$ [$10^{-3} c$]& $0.83^{+0.17}_{-0.17}$ & $0.85^{+0.18}_{-0.18}$ & $0.96^{+0.17}_{-0.15}$ & $0.83^{+0.17}_{-0.17}$ & 1.23 & $1.079_{-0.190}^{+0.197}$\\
        $RA$ [deg]& $166^{+16}_{-15}$ & $167^{+17}_{-15}$ & $173^{+12}_{-13}$ & $166^{+16}_{-15}$ & $167.94$ &  $172.95_{-11.76}^{+12.77}$\\
        $Dec$ [deg]& $10^{+19}_{-19}$ & $16^{+19}_{-18}$ & $2^{+16}_{-15}$ & $12^{+19}_{-19}$ & $-6.94$ &
        $-13.31_{-13.09}^{+15.39}$ \\
        $v_\mathrm{bulk}[10^{-3}c]$ & 0 & 0 & implicit & 0 & 0 & 
        $0.522_{-0.077}^{+0.081}$
    \end{tabular}
    \tablefoot{As default we 
use the full Pantheon sample, fit a flat $\Lambda$CDM model and include
the covariance of peculiar velocities via the matrix $V$ as described 
in the text. The label `hostz' refers to the subsample with redshift 
measurements from host spectra. `$v_e$ corrections' means we applied 
corrections from \citet{Carrick2015}. `curvature' means we fit a non-flat
$\Lambda$CDM model. `$H_0$ fixed' refers to assuming a value of 
$H_0 = 73.2$ km/s/Mpc and fixing the Solar motion to the value inferred 
from the CMB dipole. Finally, with `bulk motion' we refer to 
assuming a bulk flow as inferred by \citet{Carrick2015}.}
    \label{tb:contPant}
\end{table*}

\begin{table}
\centering
\caption{Planck best-fit values and credible intervals.}
    \begin{tabular}{c|c|c}
         &  flat & curvature \\
        \hline
        $\Delta H_0$ [km/s/Mpc] & $-5.8 \pm 0.5$ & $-9.6^{+2.1}_{-2.3}$ \\
        $\Omega _M$ & $0.315 \pm 0.007$ & $0.35\pm 0.03$ \\
        $\Omega _K$ & $0$ & $-0.044^{+0.033}_{-0.034}$\\
        $v_o$ [$10^{-3} c$]& $1.23357^{+0.0036}_{-0.0036}$ & same \\
        $RA$ [deg]& $167.942^{+0.007}_{-0.007}$ & same \\
        $Dec$ [deg]& $-6.944^{+0.007}_{-0.007}$ & same
    \end{tabular}
    \tablefoot{Planck TT+TE+EE+lowE+lensing results of \citet{PlanckVI}, 
the Solar dipole is reported from \citet{PlanckI}.}
    \label{tb:contPlanck}
\end{table}

\subsection{Nuisance parameter}

The nuisance parameter $\Delta H_0$ shows some (expected) 
degeneracy with $\Omega_M$, but none with the inferred 
proper motion of the Solar system. Conversely, this also means that 
the value of the Hubble rate measured by means of calibrated SNe cannot be 
reduced by a significant amount by assuming a cosmic rest frame that does not 
coincide with the CMB dipole frame. The effect of 
including the full covariance matrix and allowing for an arbitrary Solar velocity in the analysis is an insignificant increase $\Delta H_0 = 0.23$ km/s/Mpc. 

\subsection{Matter density}

The best-fit matter density agrees very well with the CMB analysis \citep{PlanckVI}.

\subsection{Proper motion of Solar System}

The inferred direction of the Solar system proper motion is found to be 
consistent with the value inferred from \citet{PlanckI} within one sigma. 
The velocity itself is found to be lower than the one inferred from the CMB 
dipole \citep{PlanckI}. The p-value for agreement with the CMB dipole is $0.0095$. 
The median Solar velocity is $2.4 \sigma$ below the value inferred from the 
CMB dipole \citep{PlanckI}. All subsamples of the Pantheon catalogue that 
are sensitive to $v_o$ show the same trend and we find no parameter degeneracy 
with the other cosmological parameters. This means that low redshift SNe (up to $z \sim 0.1$), 
spread out over a wide area on the sky, are an excellent tool to infer the 
cosmic rest frame independently from the CMB. The precision of that inference 
is comparable with the precision of current radio surveys \citep{Siewert2021}. 

\subsection{Host galaxy redshifts}

\begin{figure}
    \centering
    \includegraphics[width=0.49\textwidth]{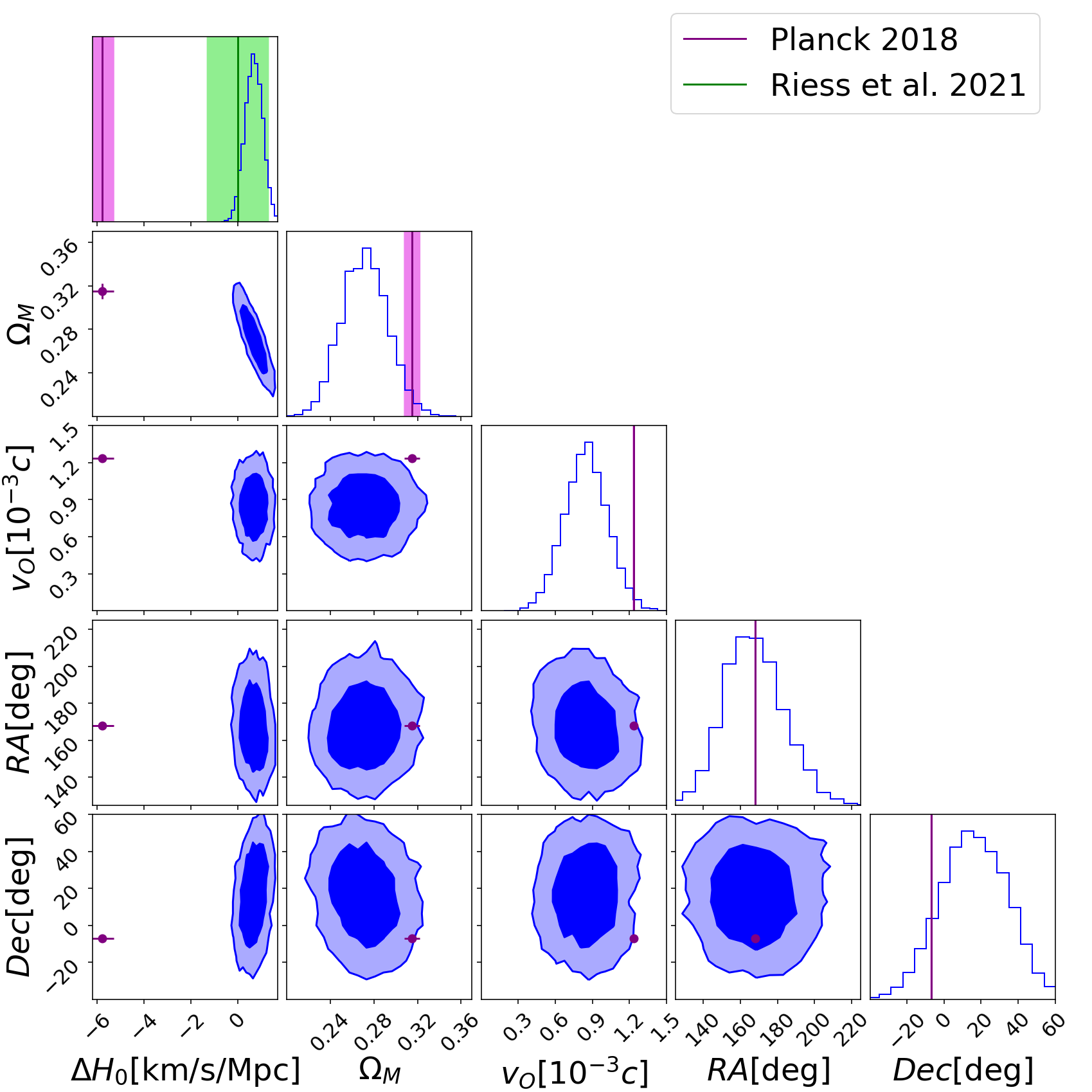}
    \caption{Posterior distributions using only SNe with host galaxy redshifts.}
    \label{fig:contHostOnly}
\end{figure}

In order to test the robustness of our results, we also restricted our
analysis to SNe with host galaxy redshifts. 
The results are shown in Fig.\ref{fig:contHostOnly}. With respect to the matter density, we 
find significant differences to the full Pantheon catalogue as already reported by \cite{Steinhardt2020}. 
In fact, $\Omega _M$ is $2.1 \sigma$ below the CMB estimate from \citet{PlanckVI}. However, 
the inferred value of $v_o$ is fully consistent with the full Pantheon sample and is $2.1 \sigma$ 
lower than the value inferred from the CMB dipole \citet{PlanckI} with a p-value of $0.015$.

\begin{figure}
    \centering
    \includegraphics[width=0.49\textwidth]{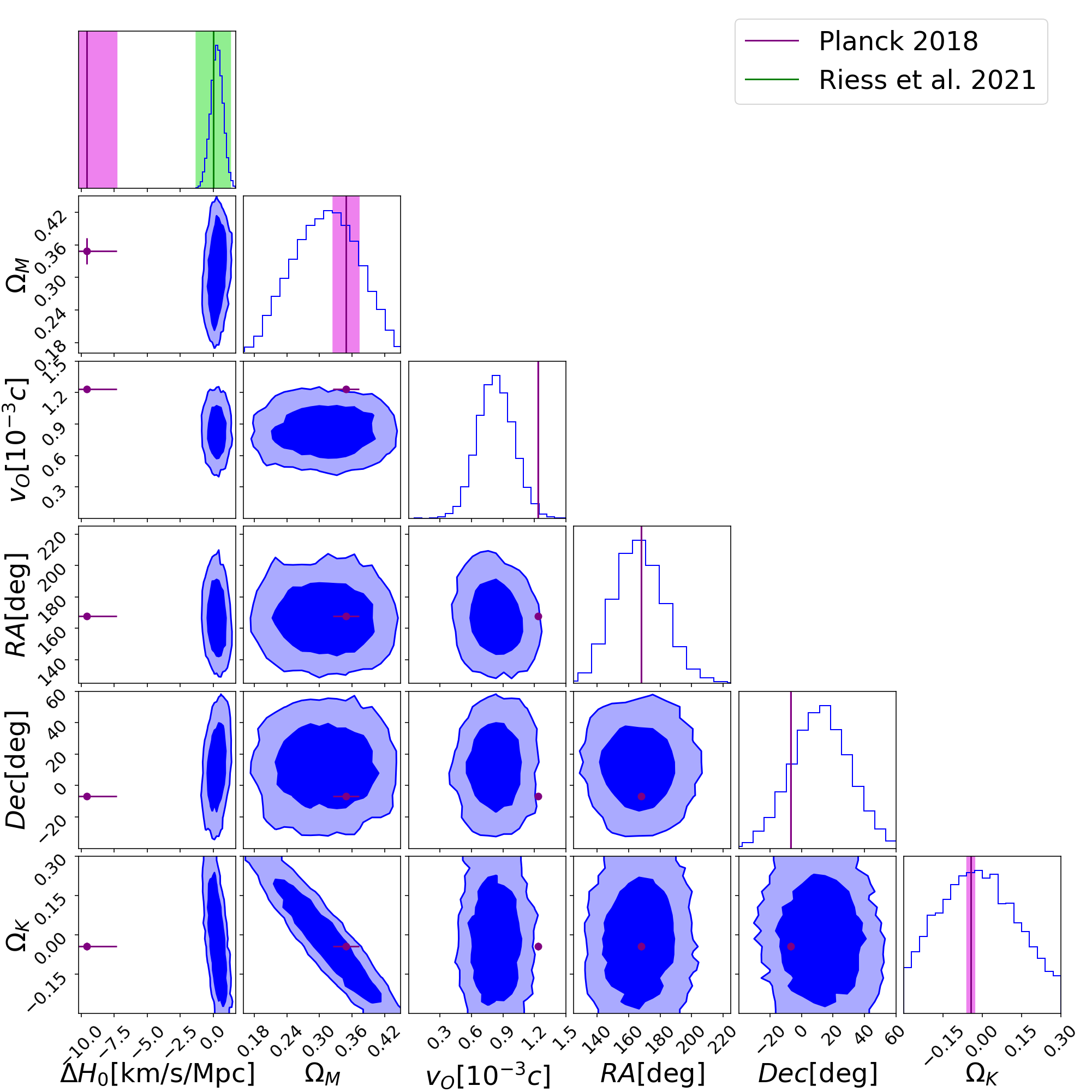}
    \caption{Posterior distributions allowing for non-zero curvature.}
    \label{fig:contCurv}
\end{figure}

\subsection{Curvature}

Allowing for curvature and introducing $\Omega _K$ as a sixth model parameter (see Fig.\ref{fig:contCurv}) leads to increased uncertainties for all varied parameters. Because $\Omega _K = 0$ is consistent with our analysis we neglect curvature in all the following analysis to save computation time and decrease the uncertainty. 
It is interesting to note that the best fit value is slightly negative, as for the extended Planck analysis \citep{PlanckVI}, but with much larger uncertainty and thus without any statistical significance. 
The inferred proper motion of the Solar system is not affected by the 
extra parameter.

\subsection{Minimal redshift \label{sec:5.6}}

\begin{figure}
    \centering
    \includegraphics[width=0.49\textwidth]{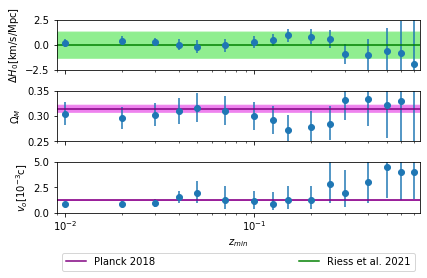}
    \caption{Best-fit parameters based on SNe with redshifts larger than $z_{min}$. All five parameters were varied ($\Delta H_0$, $\Omega _M$, $v_o$, $RA$, $Dec$).}
    \label{fig:z_min}
\end{figure}{}

For further insight we repeated the analysis omitting SNe with redshift $z_\mathrm{hel}<z_\text{min}$, in 
order to study the influence of nearby SNe. The dependence of the parameters $\Delta H_0$, 
$\Omega _M$ and $v_o$ on $z_\text{min}$ is shown in Fig.~\ref{fig:z_min}. In this analysis we allowed $v_o$ to vary in a wider range between 0 and $0.01c$. The decreasing 
number of SNe leads, as expected, to increasing uncertainties for high $z_\text{min}$. 
$\Delta H_0$ and $\Omega _M$ both show a clear difference from the full data set at $z_\text{min} > 0.1$. For $\Omega _M$ the values decrease between $z_\text{min} = 0.1$ and $z_\text{min} = 0.2$ below the Planck value and subsequently increase above it at $z_\text{min} > 0.3$. $\Delta H_0$ shows a similar evolution, increasing between  $z_\text{min}$ = 0.1 and $z_\text{min}$ = 0.2, decreasing afterwards.  At $z_\text{min} > 0.3$ the uncertainties also grow quite large due to the decreasing number of SNe. Between  $z_\text{min} = 0.2$ and $z_\text{min} = 0.3$ the number of SNe decreases from 637 to 421. SNe
The decrease of $\Delta H_0$ at higher redshifts is in line with the dependence of $H_0$ 
on the probed SN redshifts described for example by \citet{Dainotti2021}, 
who extrapolated this trend to the CMB measurements and is also seen 
in the analysis of time delays in gravitational lenses \citep{Wong2020, Millon2020}.

While the uncertainties in $\Delta H_0$ and $\Omega_M$ start to increase significantly at $z_\text{min} > 0.2$, for $v_o$ sizeable  uncertainties make the analysis less conclusive at $z_\text{min} > 0.05$. Above $z_\text{min} = 0.07$, the inferred value of $v_o$ lies within one $\sigma$ of the CMB inferred Solar velocity. The increase in $v_o$ for $z_\text{min} > 0.03$ 
could be a reflection of a large bulk flows as reported in 
\citet{Carrick2015}, when the Solar velocity is inferred from 
the CMB dipole. This is further discussed in Sect. \ref{sec:5.8}

Our findings are also consistent with the analysis of \citet{Colgain2019} who measured 
the matter density for $z_\mathrm{CMB} < z_\mathrm{max}$ from the original Pantheon catalogue.
A slight hint for a matter underdensity  at $z_\mathrm{max} < 0.2$ is reflected in our analysis by the 
increase in the matter density above $z_\mathrm{min} = 0.2$.

\subsection{Peculiar velocities}

\begin{figure}
    \centering
    \includegraphics[width=0.49\textwidth]{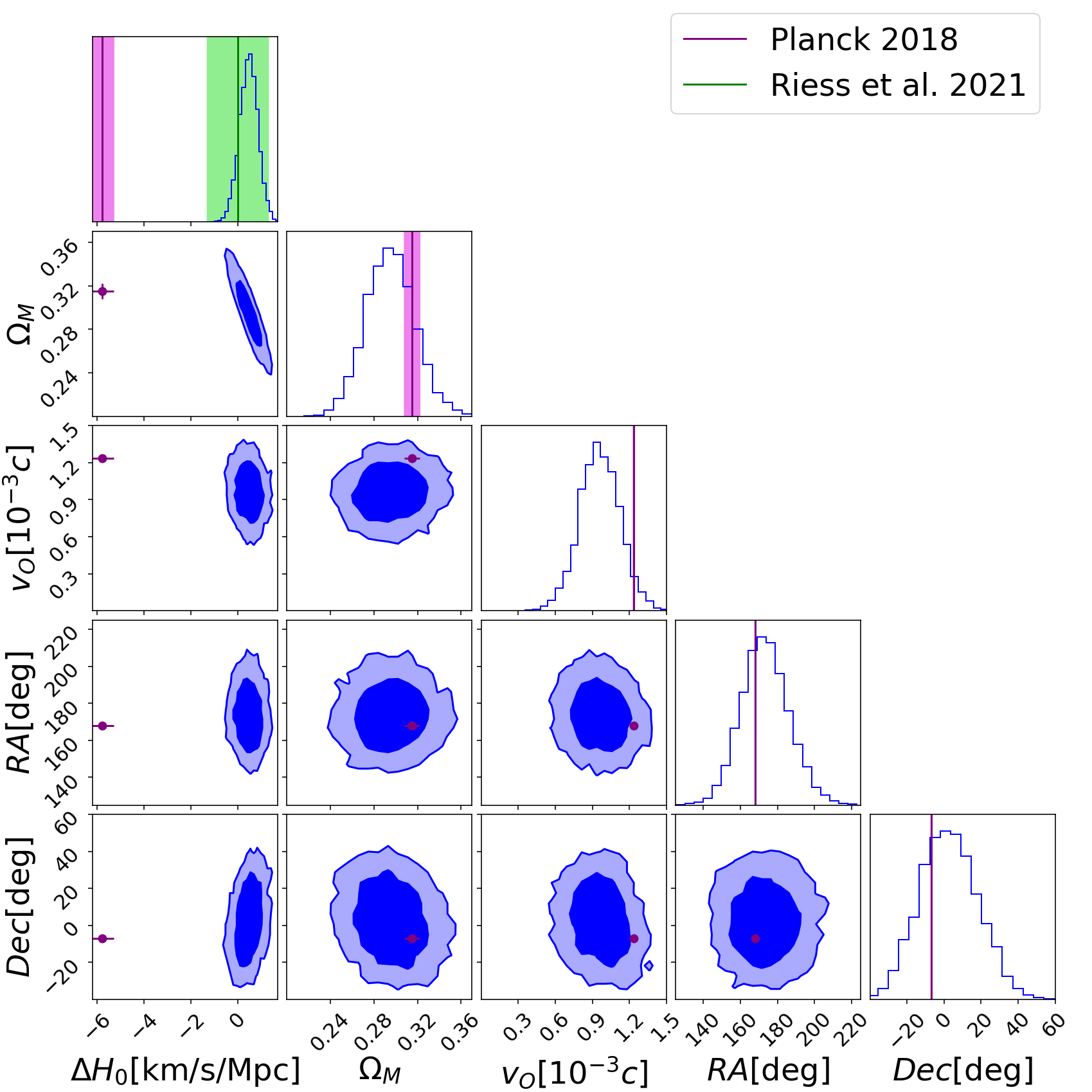}
    \caption{Posterior distribution using peculiar velocities corrections from \citet{Carrick2015}.}
    \label{fig:contvE}
\end{figure}

Up to this point we used the structure covariance $V$ to account 
for peculiar velocities of SNe. In order to test the effect of 
applying reconstructed peculiar velocities, we now use 
peculiar velocities from \citet{Carrick2015} for 
$z_\mathrm{hel} < 0.1$ instead. The resulting fits are shown in Fig.~\ref{fig:contvE}.

Correcting for peculiar velocities in this way changes the values by less than $1 \sigma $ for all varied parameters. The largest change is found in $v_o$, which increases by almost 
$1 \sigma$ and therefore reduces the tension with the 
velocity inferred from the CMB dipole to $1.7 \sigma$ and increases the p-value to $0.046$.

We conclude that our estimate 
of the proper motion of the Solar system is also robust w.r.t.\ peculiar velocity corrections. However, such corrections should be considered with great care. Firstly, the inference of peculiar velocities from observed redshifts and the observed matter density distribution relies on an assumed cosmic rest frame in the first place. Secondly, the uncertainties of the inferred peculiar velocities are hard to quantify. It is certainly inconsistent to apply corrections and to include the covariance matrix $V$. Thirdly, peculiar velocity corrections should actually be applied to host galaxy redshifts only, as the relative motion between a SN and its host cannot be known and can easily amount to a sizeable fraction of the 
peculiar velocity, see also the discussion in \citep{Peterson2021}. 

\subsection{Bulk flows}
\label{sec:5.8}

Let us finally investigate whether the inferred solar motion could be partially attributed to a bulk flow of SNe at small cosmological distances. Previous works claimed the existence of such a bulk flow \citep{Kashlinsky2008, Watkins2009, Dai2011, Feindt2013, Carrick2015}. If the Local Group's motion would be aligned with the motion of other nearby structures, one would expect to find that the Sun moves slower with respect to small redshift SNe as compared to higher redshift SNe.
Qualitatively, this is indeed the case as can be seen in Fig.~\ref{fig:z_min}. Excluding SNe at $z <0.04$ leads to larger best-fit values for $v_o$. However, if all assumptions that we have made in this analysis would be correct, then the effect of such a bulk motion should actually be included via the covariance matrix (\ref{eq:pecVelCov}). Apparently, the inferred uncertainty of the Solar velocity when including SNe in the range $0.01 < z < 0.04$ is too small in order to conclude that this is an expected effect. There are however two limitations to that argument. Firstly, it assumes that only effects of linear fluctuations are relevant, and secondly it assumes that we use the correct cosmological model.

In order to test the bulk flow hypothesis, we modify our fit by assuming that all SNe at $z < 0.03$ (motivated by the analysis presented in Sect.~\ref{sec:5.6}) have the same value of $v_e = v_\mathrm{bulk}$ in the fixed direction $RA_\mathrm{bulk} = 194$ deg and $Dec_\mathrm{bulk} = -57$ deg, as found by \cite{Carrick2015}. Using a flat prior for the additional parameter $v_\mathrm{bulk}$ does not allow us to break the degeneracy between $v_o$ and $v_\mathrm{bulk}$. In order to break that degeneracy, we added the Gaussian prior $v_\mathrm{bulk} =159 \pm 23 $km/s, taken again from \cite{Carrick2015}. The result of that fit is shown in Fig.~\ref{fig:contbulk} and table \ref{tb:contPant}. As expected, for $v_\mathrm{bulk}$ we essentially recover the prior and we find the inferred solar motion to be consistent with the one inferred from the CMB dipole. This is no surprise as \cite{Carrick2015} used the CMB dipole as an input to infer the bulk motion.

\begin{figure}
    \centering
    \includegraphics[width=0.49\textwidth]{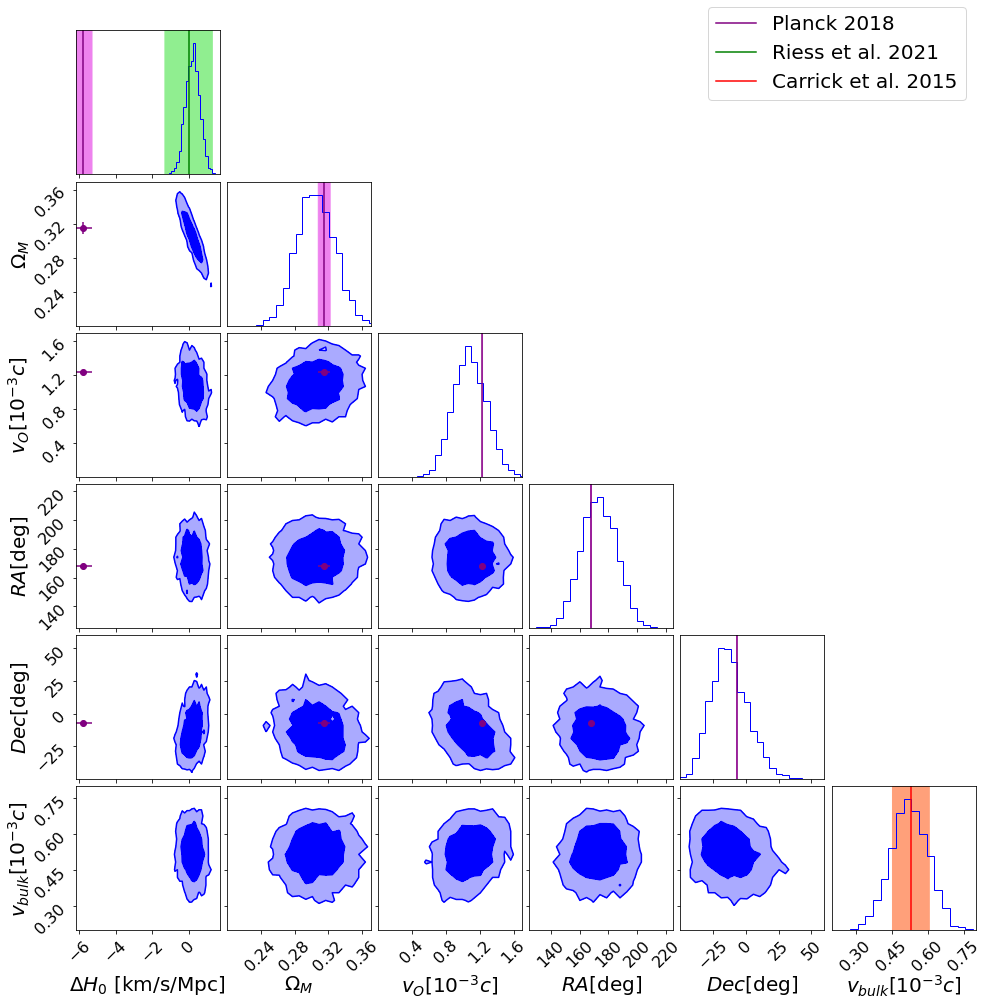}
    \caption{Posterior distribution assuming a constant bulk flow for all SNe with $z_\mathrm{hel} < 0.03$ in the fixed direction $RA_\mathrm{bulk} = 194$ deg, $Dec_\mathrm{bulk} = -57$ deg with a Gaussian prior on the bulk velocity, $v_\mathrm{bulk} =159 \pm 23 $km/s. Bulk direction and velocity prior are chosen according to \citet{Carrick2015}.}
    \label{fig:contbulk}
\end{figure}

Thus, bulk flows as discussed in the current literature could
be responsible for a smaller inferred Solar 
motion w.r.t.~a sample of SNe at small redshifts.

\section{Conclusions \label{sec:6}}

Modern cosmology describes the Universe in the context of spatially homogeneous 
and isotropic space-time, the class of the Friedmann-Lema\^itre models. These 
models have in common that a cosmic rest frame exists which is typically 
identified with the frame defined by the CMB dipole, and consequently the 
peculiar motion of the barycentre of the Solar system is inferred from the CMB. 

Here we have tested this hypothesis of a purely kinematic CMB dipole by means 
of SNe compiled in the Pantheon catalogue \citep{Scolnic2018}. 
The SN redshift-distance modulus relation, see equation (\ref{eq: mu}), 
is sensitive to the radial peculiar motions of the SNe and the radially projected
peculiar motion of the observer. While (for non-relativistic velocities) 
the SN redshift depends on $(\mathbf v_e - \mathbf v_o)\cdot \mathbf n$, 
the distance modulus depends on $(2 \mathbf v_e - \mathbf v_o)\cdot \mathbf n$, 
as has been first realised by \citet{Sasaki1987}. While inferring the peculiar 
velocities of the SN host galaxies needs additional observations and is limited 
to rather small redshifts, see e.g.\ \citet{Carrick2015}, the effect of the 
Solar system motion is coherent and affects all SNe at all redshifts, which 
allows us to actually use a direction dependent analysis of SN distance moduli 
to measure the Solar system proper motion.  

\begin{figure}
    \centering
    \includegraphics[width=0.49\textwidth]{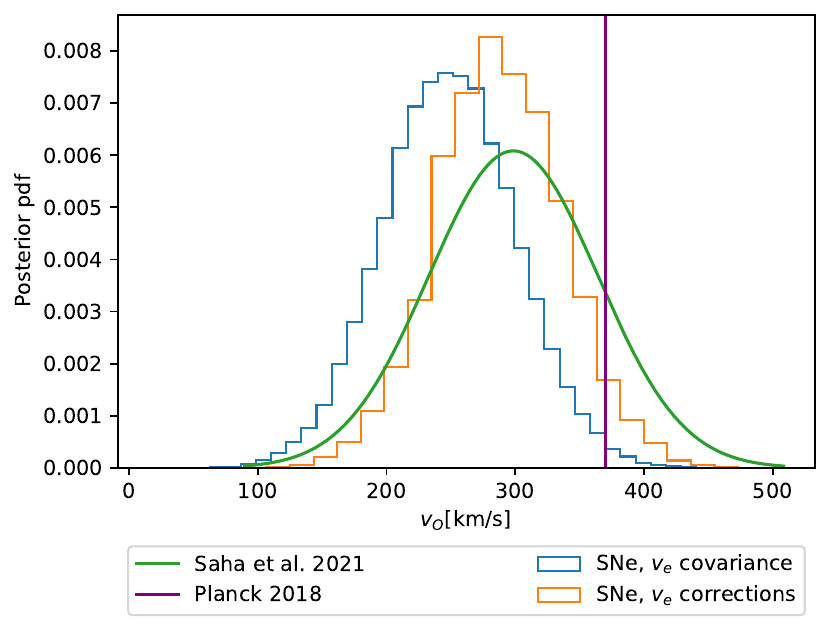}
    \caption{Comparison of the posterior distribution for different measurements of the proper motion of the Sun. We show our own results taking the emitter peculiar velocities into account by using the covariance matrix (\ref{eq:pecVelCov}) and by means of corrections according to \citet{Carrick2015}, respectively. For comparison we also show the results from the analysis of high multipole moments of the CMB \citep{Saha2021} and assuming that the CMB dipole is entirely explained by Solar motion \citep{PlanckI}. In the latter case, the uncertainties are too small to be displayed properly, instead we indicate the value by a vertical line.}
    \label{fig:summary}
\end{figure}

It has been pointed out before that the Pantheon catalogue suffers from 
inconsistencies regarding the quoted values of heliocentric redshifts 
\citep{Rameez2019, Steinhardt2020}. Indeed, the Pantheon sample was not compiled with foresight of direction dependent studies, which implies that there might be other issues that we could not identify in this study.
We used an improved version of the Pantheon catalogue (see \citet{Steinhardt2020} and App.~\ref{appendix:DataErrors})
to measure the cosmic SN rest frame and found that the direction of motion of the 
Solar system agrees well with the CMB dipole direction, but that the inferred 
Solar system velocity is well below the one inferred from the CMB dipole 
(see table \ref{tb:contPant}). Our findings are summarized in 
Fig.\ \ref{fig:summary}. The null hypothesis of a purely kinematic CMB dipole 
is found to have a p-value of $0.0095$ when accounting for the unknown 
peculiar motion of SNe by means of the covariance matrix from linear theory 
\citep{Huterer2017}. Correcting for peculiar motion \citep{Carrick2015}, which 
implies also that we must assume a cosmic rest frame, the p-value 
increases to 0.046. We therefore conclude that it is premature to reject 
the hypothesis of a purely kinematic CMB dipole, however, it is interesting to 
note that none of our tests produced a Solar velocity exceeding the one from 
the CMB dipole. 

We tested the robustness of our findings with respect to the self-consistency 
of the Pantheon catalogue and the addition of further cosmological parameters, 
and we confirmed that peculiar velocity corrections affect the final result but
move the median values by less than one sigma. As already pointed out by 
\citet{Steinhardt2020}, using host galaxy redshifts does affect the inferred 
matter density, but turns out to lead to insignificant changes in the estimate of
the Solar system proper motion. 

We have also shown that bulk flows could explain why 
the solar motion appears to be slower w.r.t.\ nearby SNe as compared to more distant ones, but the inference of such a large scale bulk flow is depending heavily on the assumption that the cosmic rest frame is defined by the CMB dipole. The task of this work is not to measure such a bulk flow, but it is clear that more data will be needed to be able to disentangle the effect of bulk flows and the solar motion, which is possible in principle when traced over a range in redshift and a wide survey area, due to the already mentioned $(2 \mathbf v_e - \mathbf v_o)\cdot \mathbf n$ dependence of the distance modulus ($v_o$ must not depend on redshift, while $v_\mathrm{bulk}$ should).

It is also interesting to compare our findings with estimates of the matter dipole 
in \citet{Secrest2021} and \citet{Siewert2021}, where an excess dipole was found 
in galaxy number counts pointing also towards the CMB dipole. That excess could 
in principle be explained by a much larger motion of the observer. While 
the frequency dependence of the matter dipole excess found in 
\citet{Siewert2021} is certainly inconsistent with such an 
explanation, 
the finding of this work is as well, indicating that the 
search for systematic issues and an unexpectedly large contribution from a local clustering dipole must continue.

Another attempt 
to measure the Solar proper motion from SN data was presented 
by \citet{Singal2021}, based on a different method, using the 
JLA sample \citep{jla} and the uncorrected Pantheon sample 
\citep{Scolnic2018}. Their estimate for the direction 
agrees with ours and shows comparable, yet slightly larger, 
uncertainty. Contrary to us, they infer a Solar velocity of about 
four times larger than inferred from the CMB dipole. On the one 
hand, the factor of four would be in line with the findings 
on the matter dipole for quasars \citep{Secrest2021}, on the 
other hand it is in stark disagreement to this 
work and to the higher CMB multipole moments \citep{Saha2021}. 
The reason for this strong disagreement is unclear to us, it 
might be related to inconsistent heliocentric redshifts as 
discussed in the appendix of \cite{Steinhardt2020}.

The recently published Pantheon+ data set \citep{Brownsberger2021, Peterson2021, Carr2021, Scolnic2021} 
will allow to improve on the analysis presented here. 
The new data set contains a total of 1800 SNe and the number 
of small-$z$ SNe, which are of particular value for our analysis, 
are tripled (585 with $z < 0.08$ compared to 194 in the Pantheon catalog). 
The statistical power of such a larger sample should reduce the credible intervals 
obtained in this work 
by a factor of roughly $1/\sqrt{3}$. If we assume that the best-fit values would 
not change with the new data, the larger statistics should be enough to test 
the kinematic dipole hypothesis on a statistically significant level. 
A recent analysis of this dataset \citep{Brout2022} points 
to a dipole in the redshifts, pointing roughly opposite to the 
CMB-dipole. In contrast to our work, they do not attempt to infer 
the cosmic rest-frame from the SNe.

We conclude that SNe should be used to establish the cosmic rest frame 
independently from the CMB and without making assumptions on the 
rest frame itself. We think it is best to start from the 
heliocentric redshifts and magnitudes and to account for all 
unknowns in the covariances. We have demonstrated that this 
approach preserves the statistical power to constrain 
the matter density and curvature and we hope that we can encourage to put more emphasis on direction dependent effects in observational studies of the expansion of the Universe.

\begin{acknowledgements} 

The authors would like to thank Daniel M.\ Scolnic, Charles L.\ Steinhardt and Albert Sneppen for 
providing their data and giving insights into their work. We also thank Eoin {\'O} Colg{\'a}in for insightful 
comments and Gaurav Kumar for useful discussions regarding the data analysis. We also thank the anonymous referee for their excellent questions, suggestions, and comments.
The results of this paper have been accomplished using {\tt NumPy} \citep{numpy}, 
{\tt SciPy} \citep{scipy}, {\tt matplotlib} \citep{matplotlib}, {\tt lmfit} \citep{lmfit}, {\tt emcee} 
\citep{emcee2013}, {\tt corner} \citep{corner2016} and {\tt CAMB}\footnote{\url{camb.info/}} \citep{camb}.

This research made use of {\tt Astropy},\footnote{\url{www.astropy.org/}} a community-developed core Python package for Astronomy \citep{astropy:2013, astropy:2018}. 

\end{acknowledgements}

\bibliographystyle{aa} 
\bibliography{references}

\begin{appendix}

\section{SN positions}
\label{appendix:DataErrors}

For 18 SNe from the GOODS and SCP surveys we were not able to reproduce 
the redshift conversion between heliocentric and CMB frames given is  
\citet{Steinhardt2020}. For two of the SNe in question the 
declination given by \citet{Steinhardt2020} had a sign error and for 
one neither right ascension nor declination were given. 
The corrections are listed in table \ref{tab:PantheonCorrections}.

It seems that \citet{Steinhardt2020} assumed the same value of 
$\cos{\theta}$ for all 18 GOODS and SCP SNe, while the observations 
were taken in four different regions of the sky. 
We conclude that the values of $z_\mathrm{CMB}$ reported by \citet{Steinhardt2020} 
should be updated in the improved catalogue, applying correct SN 
positions. We did so in our analysis. 

\begin{table}
    \centering
    \caption{Corrected SNe positions.
    \label{tab:PantheonCorrections}}

    \begin{tabular}{c|c|c|c|c}
       SN  & $RA_{\text{Steinhardt}}$ & $Dec_{\text{Steinhardt}}$ & $RA_{\text{cor}}$ & $Dec_{\text{cor}}$ \\
       \hline
        Eagle & 189.336 & -62.228 & 189.336 & 62.228\\
        Frodo & 0 & 0 &53.093 & -27.740 \\
        Ombo & 53.106 & 27.751 & 53.106 & -27.751
    \end{tabular}
    \tablefoot{SNe with wrong positions in the \citet{Steinhardt2020} 
    catalogue and the corrected values from \citet{Riess2007} and \citet{Riess2004}}
\end{table}

\section{Subsamples}
\label{appendix:SubSamples}

Here we present one- and two-dimensional posterior distributions for the 
individual surveys included in the Pantheon data set (Figs.~\ref{fig:B1} -- 
\ref{fig:B2}) and combinations of pairs of surveys (Figs.~\ref{fig:B3} -- 
\ref{fig:B4}). The combination of two surveys tightens the constrains in 
all cases (as expected) and combinations that include the SNLS sample tend 
to prefer smaller values of $\Omega_M$, but not at a statistically 
significant level and not when combined with the PS1 sample. 
Furthermore, we compared combinations that omit a single survey 
(Figs.~\ref{fig:B5}), which are also consistent with each 
other. However, we would like to point out that omission of the PS1 survey 
leads to a slightly smaller value of $\Omega_M$, as compared with the 
best-fit from the CMB.

Further inspection of the posteriors of individual surveys reveals 
that surveys with mean redshift above $0.3$ have very little 
constraining power on the proper motion of the Solar system 
but are good in constraining the matter density. We also see that 
removing the CfA surveys leads to a significant increase in the 
uncertainty of the Solar system proper motion. In order to obtain 
meaningful constraints on $v_o, RA, Dec$, at least two surveys must
be combined. All combinations are consistent with the common fit 
and therefore, we can conclude that the combination of the various 
SN samples is at least self-consistent.

\begin{figure}
    \centering
    \includegraphics[width=0.49\textwidth]{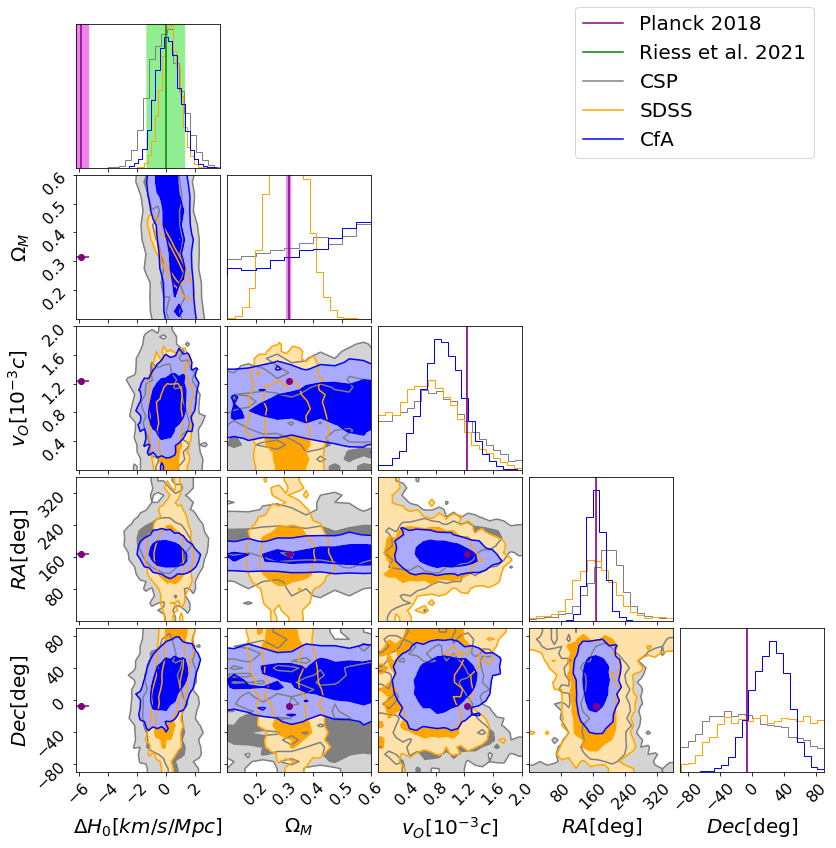}    
    \caption{Two-dimensional and one-dimensional posterior distributions for four cosmological parameters and the nuisance parameter $\Delta H_0$ using only data from CSP, SDSS and CfA, respectively. The contours show the 68\% and 95\% credibility levels.}
    \label{fig:B1}
\end{figure}

\begin{figure}
    \centering
    \includegraphics[width=0.49\textwidth]{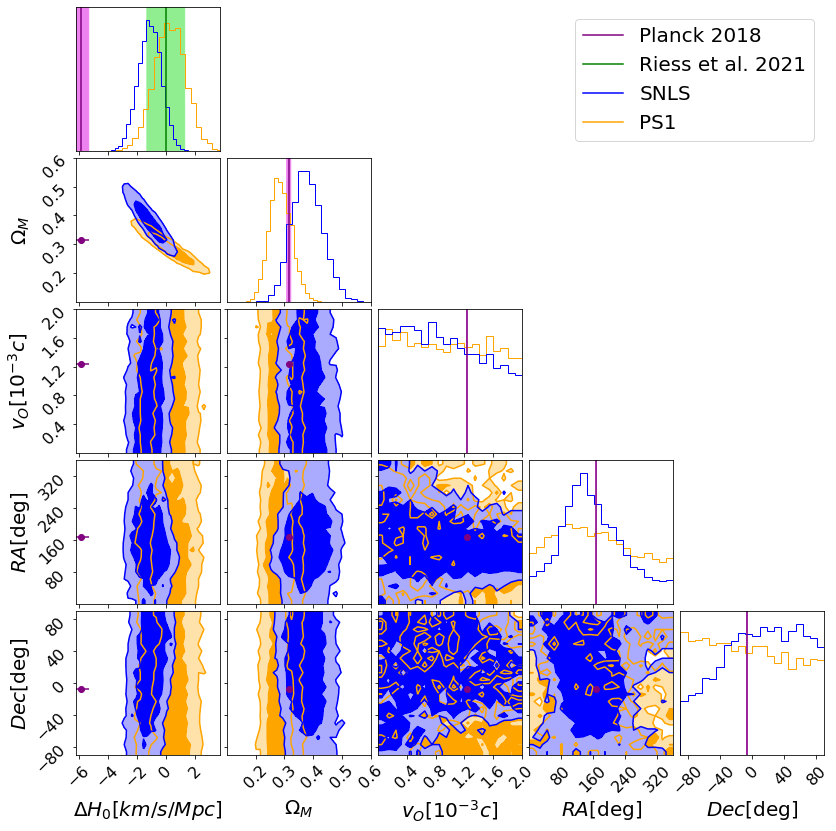}   
    \caption{As Fig.~\ref{fig:B1}, but using only data from SNLS and PS1, respectively.}
    \label{fig:B2}
\end{figure}

\begin{figure}
    \centering
    \includegraphics[width=0.49\textwidth]{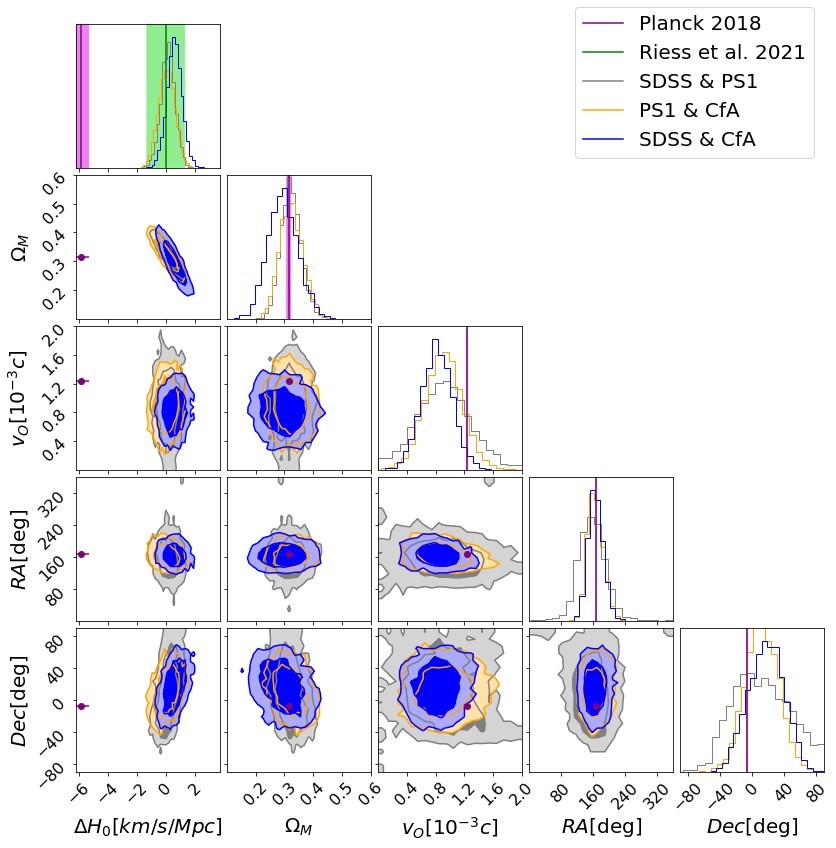}    
    \caption{As Fig.~\ref{fig:B1}, but using data from the combinations SDSS \& PS1, PS1 \& CfA and SDSS \& CfA, respectively.}
    \label{fig:B3}
\end{figure}

\begin{figure}
    \centering
    \includegraphics[width=0.49\textwidth]{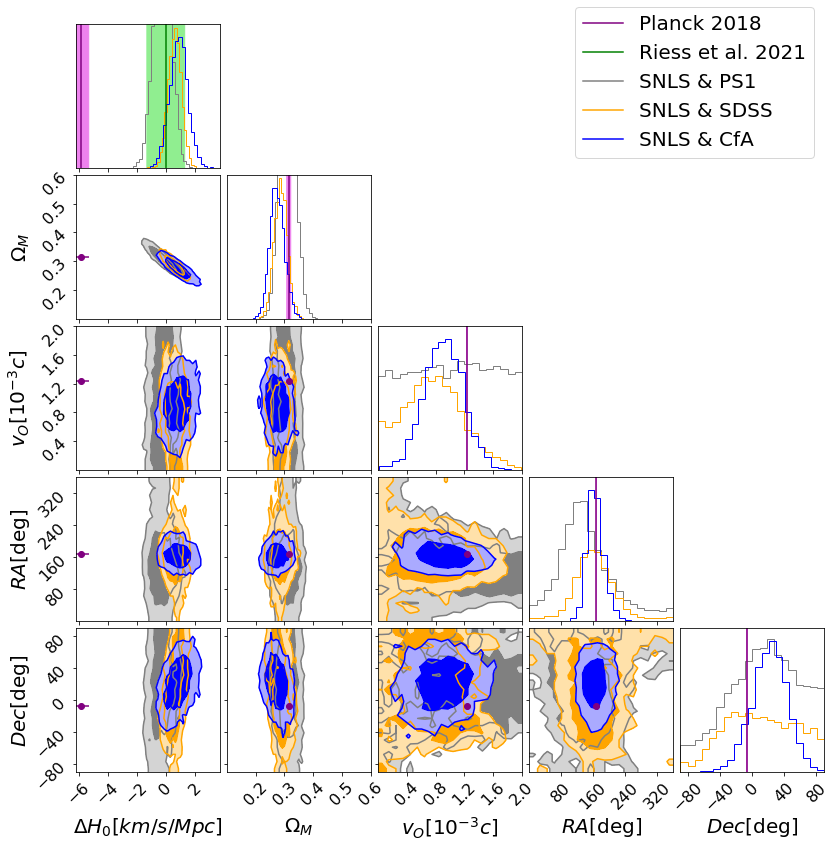}    
    \caption{As Fig.~\ref{fig:B1}, but using data from the combinations SNLS \& PS1, SNLS \& SDSS and SNLS \& CfA, respectively.}
    \label{fig:B4}
\end{figure}

\begin{figure}
    \centering
    \includegraphics[width=0.49\textwidth]{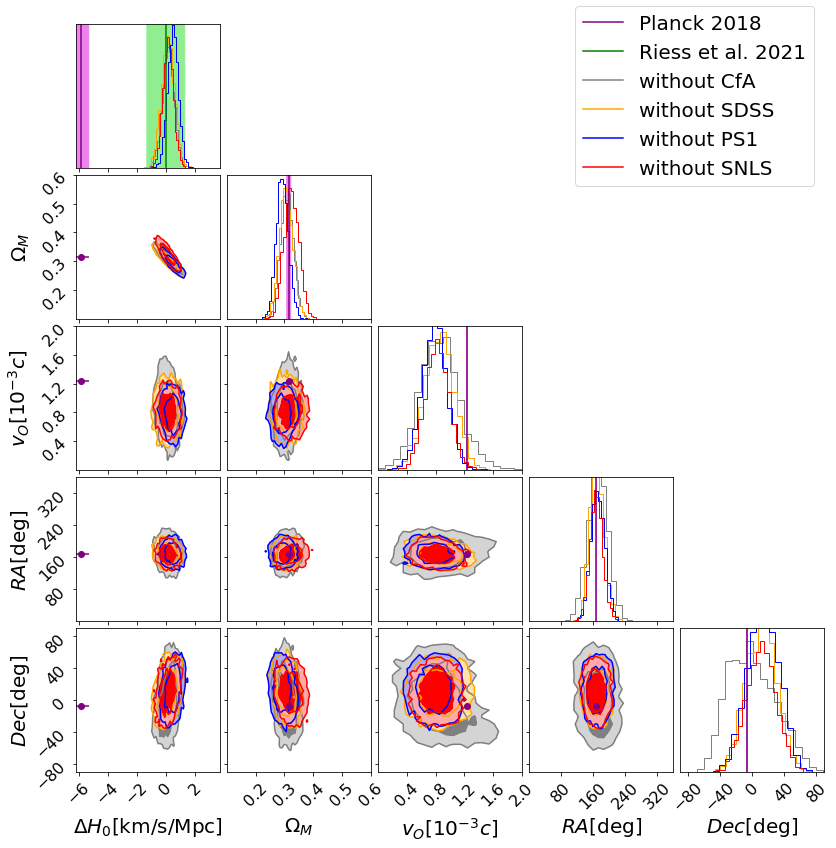}    
    \caption{As Fig.~\ref{fig:B1}, but using data from all surveys, excluding CfA, SDSS, PS1 and SNLS, respectively.}
    \label{fig:B5}
\end{figure}

\end{appendix}
\end{document}